\documentclass[referee]{mn2e}
\usepackage{graphicx}
\usepackage{journals}

\title[Fine structure above a light bridge in the transition region and corona]{Fine structure above a light bridge in the transition region and corona}
\author[L. Bharti]{L. Bharti$^{1}$\thanks{E-mail:
lokesh$\_$bharti@yahoo.co.in}\\
$^{1}$Bal Shiksha Sadan Society, Sajjan Nagar, Udaipur, Rajasthan, India\\}
\begin{document}

\date{\bf Accepted ..... . Received .......; in original form ........}

\pagerange{\pageref{firstpage}--\pageref{lastpage}} \pubyear{2013}

\maketitle

\label{firstpage}

\begin{abstract}
We present the results of multi wavelength, co-spatial and near
co-temporal observations of jets above a sunspot light bridge.  The
data were obtained with the Solar Optical Telescope (SOT) on board
Hinode, the Interface Region Spectrograph (IRIS) and the Atmospheric
Imaging Assembly (AIA) on board the Solar Dynamic Observatory
(SDO). Most of the jets in the Ca II H images show decreasing
brightness with height while in the IRIS slit jaw images at 1330 \AA~
jets show a bright leading edge. These jets show rising and falling
motion as evident from the parabolic profile obtained from the
time-distance diagram. The rising and falling speeds of the jets are
similar. These jets show a coordinated behaviour between
neighbouring jets moving jointly up and down. Some of the
jets show a plasma ejection from the leading edge which is also hotter
at the transition region (TR) and coronal temperatures
. A Similar behaviour is seen in the AIA wave bands that
suggests that jets above the LB reach up to the lower corona and
the leading edges are heated up to coronal temperatures.  Such jets
are important means of transfer mass and energy to the transition
region and corona above sunspots.

\end{abstract}


\section{Introduction}

Jet-like transients are assumed to be means of energy and mass
transport from the solar surface to the upper layers of the solar
atmosphere (Shibata et al. 2007, Morton et al. 2012).  High resolution
observations from the Hinode/SOT have revealed numerous chromospheric
anemone jets in the solar chromosphere outside active regions (Shibata
et al. 2007). These jets have typical $\lambda$-shapes which supports
the notion of reconnection between an emerging magnetic bipole and a
preexisting uniform vertical field (Yokoyama \& Shibata 1995) and
gives indirect evidence of small-scale ubiquitous low-altitude
reconnection in the solar atmosphere. There are also transient events
of cool plasma such as spicules, macrospicules and surges (Beckers
1972, Koutchmy \& Stellmacher 1976, Yamauchi et al. 2005) observed in
the chromosphere.  Spicules are observed {\bf at} the limb in the
chromospheric emission lines as well as in the UV and EUV
wavelengths. From Hinode observations of de Pontieu et al. (2007a)
the existence of two types of spicules, type-I
and type-II, is identified.  Both types of spicules are
anchored to the photosphere. However, the driving mechanisms are
different for both. Type-I spicules follow parabolic path as
seen in the time-distance diagrams, suggesting that ejected mass
returns to the solar surface while type-II spicules show
accelerations and do not return to the surface at chromospheric temperatures.
Pereira et al. (2014) show that they do return to the surface, but at transition region
temperatures. Rouppe vander Voort et al. (2009) reported on the existence of type-II spicules
on the solar disk.

Dynamic jet-like events are also observed above sunspots. Using limb
observations from Hinode in the Ca II H band and TRACE at 1550 \AA~ and
1600 \AA~ Morton (2012) reported on jets at the edge of a sunspot
which rises the chromospheric plasma to the low corona through the
transition region (TR). The leading edges of jets show enhanced
emission due to shock heating. Jet-like events above sunspot penumbrae
'penumbral micro jets' were reported by Katsukawa et
al. (2007). Penumbral microjets are bright transients observed at the
sides of penumbral filaments. Their apparent speed in the image plane
is about 100  km/s and their lifetime are around one minute. It has been
suggested that these jets are aligned with the more inclined
penumbral field (Jur\v c\'ak \& Katsukawa 2008) and caused
by reconnection between the more inclined penumbral field and
background field. There are evidences of jet-like events in the
sunspot umbra as well. Bharti et al. (2013) reported on the existence
of umbral microjets. These jets are also aligned with the background
magnetic field. These sub-arcsecond structures last for about one
minute. Small-scale, jet-like features with periodic appearance in the
chromosphere above sunspots has been found by Rouppe van der Voort \&
de la Cruz Rodr\'iguez (2013).  They originate due to the leakage of
long-period waves into the chromosphere along inclined magnetic
field. Yurchyshyn et al. (2013) described cool cone-shaped jet-like
structures in the chromosphere above the strongest magnetic part of the
umbra. They show upward and downward oscillatory motion. According to these
authors the driving mechanism of the jets are photospheric
oscillations that generate upward moving shocks.

Brightness enhancement, ejection and surge activity above light
bridges in the sunspot chromosphere have been also reported (Roy 1973,
Asai et al. 2001, Bharti et al. 2007, Louis et al. 2008, 2009, Shimizu
et al. 2009, Shimizu 2011).  Shimizu et al. (2009) and Shimizu (2011)
reported on a bright plasma ejection that was intermittent and
recurrent for more than a day above a light bridge.  Berger \&
Berdyugina (2003) reported on brightness enhancements above a light
bridge in the transition region. Louis et al. (2014) suggested that
jets above light bridges are guided by sunspot magnetic fields. These
jets are caused by magnetic reconnection between the more inclined
magnetic field in the LB and the more vertical umbral magnetic
field. The presence of opposite polarity field close to these
transient events is being hinted at by Bharti el a. (2007) and Louis
et al. (2014).

\section[]{Observations}

The largest spot (NOAA active region 2192) of the current solar cycle
was observed by IRIS, AIA/SDO and SOT/Hinode onboard on October 25,
2014. In the present study observations from
these instruments from 05:17 UT to 06:30 UT are used. Slit-jaw images
obtained by IRIS (De Pontieu et al. 2014) at 1330 \AA~(C II) with AN
image scale OF 0.166"/pixel and 7.3 sec cadence were obtained. The
SDO/AIA (Lemen et al. 2012) observations have a plate scale of
0.6"/pixel and were taken with a cadence of
12 sec. G-band and Ca II H images obtained by
Hinode (Kosugi et al. 2007) have an Image scale of 0.22"/pixel. Only a few G-band images are
available. However, Ca II H images were taken every
minute. The standard routine $'fg\_prep.pro'$ from solarsoft was
applied to the Hinode data for flat field and dark current
corrections.  We used a Wiener filter to correct for the point spread
function of the Hinode/SOT telescope.  Calibrated Level 2 data from
IRIS that include flatfield, darkcurrent and geometric corrections
are used. Similarly, level 1.5 data from SDO/AIA are used. Hinode and
AIA images were upscaled to the pixel scale of IRIS. The location of
the spot on solar disk was at $x=319\arcsec$ and $y=-328\arcsec$
which corresponds to a heliocentric angle $\theta=29^\circ$
($\mu$=0.88). The Ca II H and C II filter samples emission at $10{^4}$ and $\sim$$3\times10{^4}$ K
respectively. The AIA EUV channels cover the temperature range from $6\times10{^5}$ K to
$2\times10{^7}$ K (Lemen et al. 2012).

\begin{figure*}
\vspace{-14mm}
\centering
\includegraphics[width=200mm,angle=0]{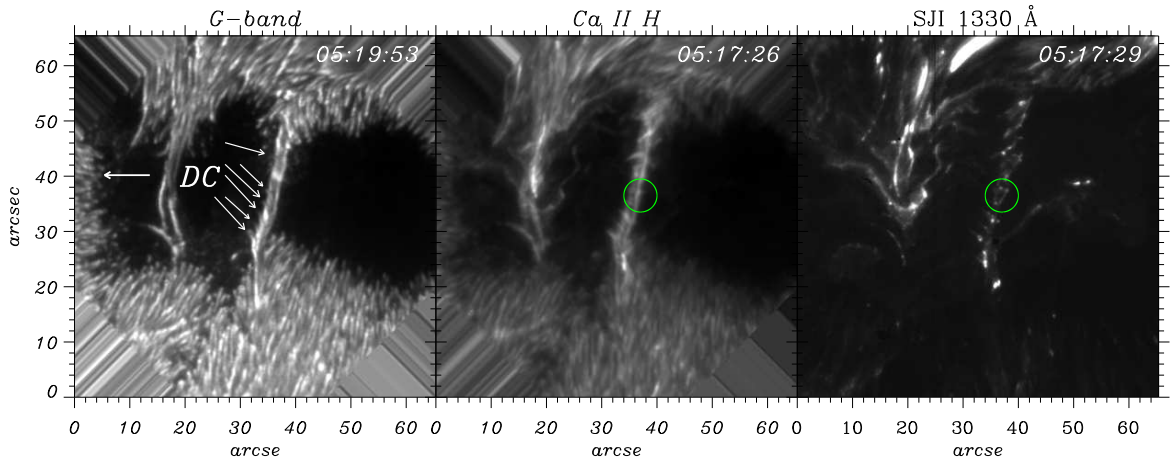}
\vspace{5mm}
\caption{Left panel: G-band image of the observed sunspot at 05:19:53
  UT.  The disk center is indicated by the arrow labelled 'DC'. The
  arrows on the left side of the right LB indicates threads in the
  LB. Middle panel: co-spatial and near co-temporal Ca II H image at
  05:17:26 UT. Right panel: IRIS slit-jaw image AT 1330 \AA~ at
  05:17:29 UT. The contrast has been enhanced in all images
  to show threads in G-band and brightening above them in
  Ca II H and IRIS slit-jaw images. Green circle indicates jets
  discussed in the main text. }
\end{figure*}

\begin{figure*}
\vspace{-215mm}

\centering
\includegraphics[width=520mm,angle=0]{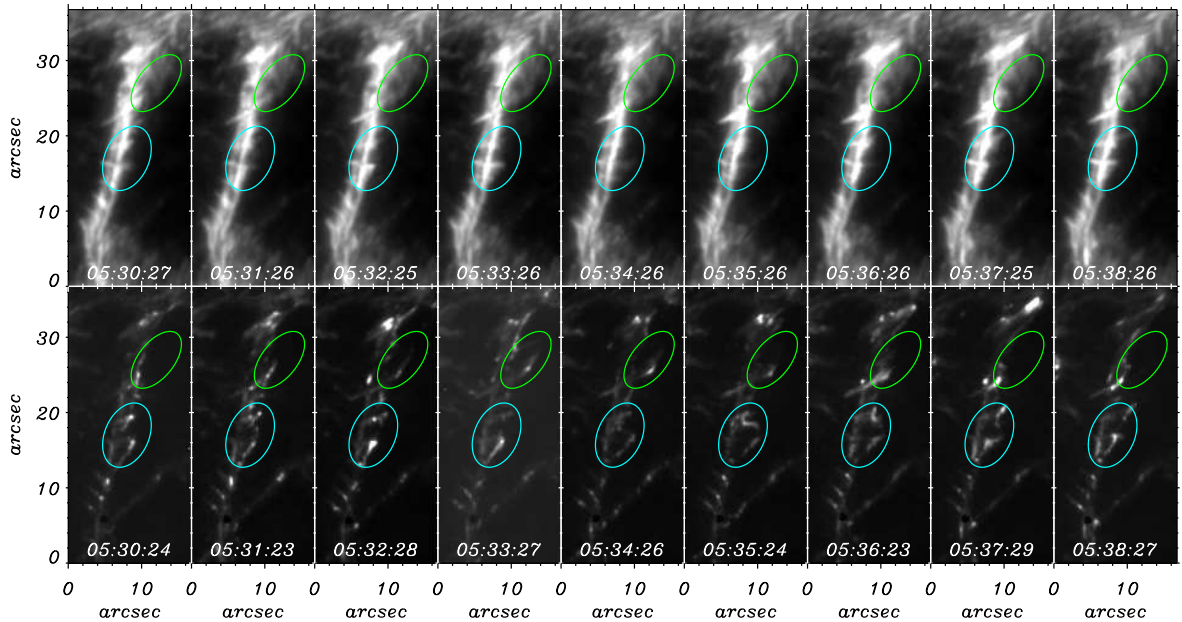}
\caption{Evolution and interaction of jets above the LB. Upper row: Ca\,
  II H images; lower row: co-spatial and near simultaneous slit-jaw
  images at 1330 \AA. The jet in the green ellipse rise and fall
  clearly. Merging and splitting of jets can be recognised in the cyan
  ellipse (see movie-II for more details which is available as on-line
  material).}
\end{figure*}

\begin{figure}
\vspace{-140mm}

\centering
\includegraphics[width=270mm,angle=0]{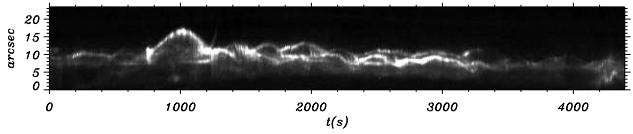}
\caption{Time-distance diagram of a stationary slit for the whole span of
  observations in 1330 \AA. The slit was oriented along the jet length. The location of
  this jet is indicated by the green ellipse in Figure 2 that achieves
  maximum height around 1000 sec. The parabolic profiles of jets
  indicates that ejected jet mass returns.}
\end{figure}

\begin{figure}
\vspace{-250mm}

\centering
\includegraphics[width=520mm,angle=0]{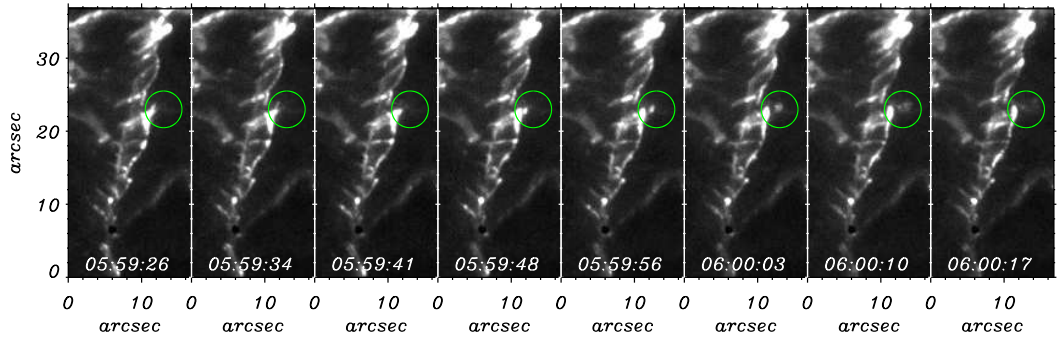}
\vspace{-10mm}
\caption{Evolution and migration of a bright blob as a result of
  interaction of jets, seen in the IRIS slit-jaw images (see Movie II
  for more details which is available as online material). The
  contrast of the images has been enhanced to see the fading of the
  bright ejection in the umbral background.}
\end{figure}

\begin{figure}
\vspace{-270mm}

\includegraphics[width=540mm,angle=0]{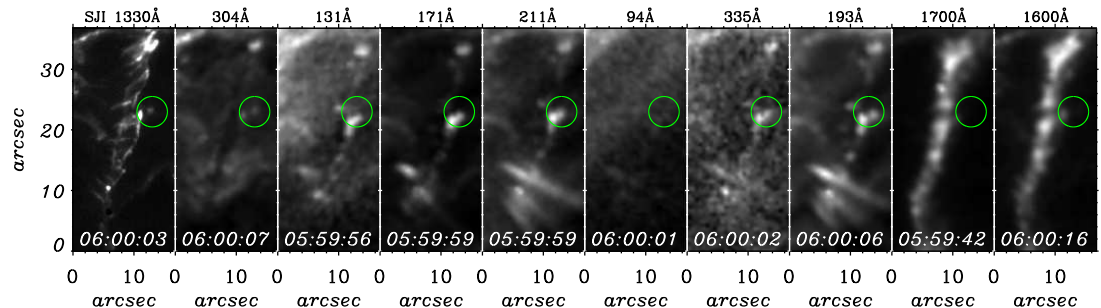}
\caption{Co-spatial and near co-temporal images of the LB in various
  wave bands. The leftmost image shows the IRIS slit-jaw image at
  1330 \AA, after that AIA images in various wave bands are
  displayed. The green circle indicates the location of the bright
  ejection shown in Figure 4. The ejected mass as well as the leading
  edge of jets are also bright at coronal temperatures.}
\end{figure}

\section{Analysis and results}

The left panel of Figure 1 shows a G-band image of the sunspot. This
spot has two filamentary light bridges (LBs). The left one is broader
than the right one. The central dark lane is more prominent in the
broader one.  We analyse the right LB in this study. There are several
thread-like structures on the left side of this LB. The locations of
these structures are indicated by arrows. These structures also have
central dark lanes, similar to penumbral filaments and
LBs. The end points lie in the dark umbra.  A co-spatial Ca
II H image from about two and half minutes earlier than the
G-band image is displayed in the middle panel. The right LB shows
bright jet-like structures at both the left and right sides. These
brightening can be seen in movie-I, available in the online
version.  The thread-like structures in the G-band image also
appear brighter in the Ca II H images and there is always enhanced
brightening on the left side of the LB. Since disk center is towards
the left side of the LB, jet-like structures in the Ca II H
images on the right side of the LB are more distinct in the dark
umbral background. The right panel illustrates a co-spatial and near
simultaneous IRIS slit-jaw image (SJI) at 1330 \AA~. Thread-like
structures are also bright in SJI but some of them have bright plasma
blobs. These thread-like structures have been chosen as a reference to
co-align the Hinode and SIJ images. Close inspection of movie-II which
is available as online material suggests that these blobs indicate
apparent loop-like shapes which rise and fall
back down. Limited spatial resolution hinders identifying
further details. The most notable features of SJI is that the jets
have bright leading edges on the right side of the LB. Some jets show
only a bright base and the bright leading edge and a void
in between. On the other hand some jets show hazy regions in
between. Examining these jets in the Ca II H image shows only a few hazy structures.
A right triangular-like structure can be seen at $x=38\arcsec$ and
$y=38\arcsec$ in the SJI image (indicated by a green circle) where
opposite is one distinct jet and hypotenuse show the bright leading
edges. In the Ca II H image only hazy opposite is hinted.  The
leading edges of the jets form a chain like structure in the SJI
images. To our knowledge such a behaviour of jets above a LB in
TR lines have not been reported earlier. We cannot rule out the
possibility that there are distinct jets (Louis et
al. 2014) and due to the lower resolution of the Ca II H data they are
not visible. Moreover, due to the limited spatial resolution of IRIS,
the bright leading edge of each jet is not clearly visible even in the
SJI. Only a chain of leading edge is visible in the
SJI.

Figure 2 demonstrate the evolution of jets in both the 1330 \AA~ and
Ca II H lines. All images are spatially co-aligned and near
co-temporal. We first discuss the rise and fall of a jet whose
location is shown by green ellipse in both wave bands. At 5:30
UT two chain-like structures can be seen in the 1330 \AA~ image at the
left side of the ellipse. The lower chain is brighter than the upper
one while the lower part of the lower chain is connected to the base
of the LB. The less brighter chain is isolated i.e. only leading edges
can be seen. In the Ca II H image these
structure appear differently. Both structures are distinct and
jet-like. The bright chain is still brighter in the Ca II
H. Both have bright bases but the brightness decreases with
height for the upper jet. With time the length of both jets increases
and they appear like a loop in the 1330 \AA~ band while
in the Ca II H band, instead of two distinct jets, they form a hazy
blob.  At 5:34 UT the jet approaches to the maximum height and then
starts to fall. Notice that there is still a hazy blob in the Ca II
H. The apparent maximum length of the apparent loops is about 2100
km. Interaction of jets in the terms of merging is indicated by a cyan
ellipse.  Initially, at 5:30 UT two distinct chains of leading edges
are seen in the 1330 \AA~ image. The lower chain is broader and
longer. These chains appear very different in Ca II H. There are two
distinct jets can be seen from which the upper one is
brighter. With time the jets rise and merge with each other. At
5:36 UT they show a one chain system in 1330 \AA~. After that
they again show two chain systems from which the upper one rises
again. The notable difference is that these jets appears only as
hazy blob structures and at some instance as jet-like in the Ca II H.

Figure 3 demonstrates a space-time diagram of a jet. This jet is
labeled by the green ellipse in Figure 2. A stationary slit is put
along the jet length. The slit is put for the total time span so that
rising and falling speeds as well as trajectories of other jets can
be identified. The jet approaches to the maximum height around
T=1000 sec. The projected rising speed of this jet is $\approx$12 km/s
and the falling speed is $\approx$13 km/s. Thus the rising and falling
speeds are similar. A similar behaviour can be seen for
other jets too. Apart from the striking rising and falling
motion, jets reoccur around the same location over the full
observational period and a coordinated behavior
of neighboring jets moving up and down can be detected. Such behaviour of jets (dynamic fibrils) have been
reported in sunspots (Rouppe van der Voort \& de la Cruz Rodr\'iguez
2013) and plage (Hansteen et al. 2006, De Pontieu et al. 2007) in the
chromospheric lines.

Figure 4 illustrates the mass ejection from a jet. In movie-II
one can see that before 5:59:26 UT there were two different leading
edges of jets.  The lower jet migrates towards the upper one and
starts to merge with it. At 5:59:34 UT the merging completes and later
on a plasma blob detaches from it. The blob moves away from leading
edge and after 6:00:17 UT it fades away. Thus the lower rising jet
pushes a pre-existing jet, destabilizes it and an ejection occurs.
This event can be seen only in the 1330 \AA~ images. Such finer
details at high spatial and temporal scale is only possible due to
IRIS capability.

\subsection{Relation to photospheric flow}

The Ca II H as well as 1330 \AA~ movies suggest that bright jets
migrate from the center of the LB towards both sides to the nearby
penumbra . A G-band movie with an uneven
cadence and only a few images as well as a HMI continuum movie (not
shown here) also show such migration of mass from the center of the LB
towards the nearby penumbra . Such a flow
pattern and its relation with chromospheric brightening is
also found by Bharti et al. (2015a) and Louis et al. (2008,
2014). Also the orientation of the jets towards the penumbra suggests
that these jets follow magnetic field lines which are more inclined
in the in the penumbra (Louis et al. 2014).

\subsection{Correlation with AIA wavelengths}

The same LB is also observed in the AIA wavebands. Since the observed
jets above LB in the 1330 \AA~ are different or not resolved in the
Hinode Ca II H images, it is imperative to see their appearance in
AIA channels. The leading edges seen in the SJI also
appear brighter than the umbral background in the AIA images. They are used as
a reference for co-alignment in an approximate manner. The plasma
ejection event displayed in Figure 4 at 6:00:03 UT is shown in Figure
5 in various wavebands. In the 304 \AA~ image all jet-like structures
appear darker from the bottom to top, only the leading edges appear
bright in the 304 \AA~. The plasma ejection event (bright leading edge
and detached plasma blob) indicated by a green circle is also
brighter. Similarly, the leading edge of the jet at $x=12\arcsec$ and
$y=34\arcsec$ is also bright while the lower part of the jet is still
dark. A movie of the 304 \AA~ images (movie-III IS available as
online material) show dark material falling down above LB at
$x=13\arcsec$ and $y=12\arcsec$ from 5:23:07 UT to 5:32:56 UT. In the
131 \AA~ image only the leading edges of the jets appear
brighter. Since the 131 \AA~ image has been obtained approximately
7 sec earlier, the location of ejection event
is shifted. Note that there is a shift also due to the projection effect of formation height of
different wavebands. The structuring in the 171 \AA~ and 211 \AA~ images are
similar as only the leading edges are visible. Some bright loops from
the umbral core are also visible above the LB
structure. There are no signatures of jets in the 94 \AA~
image. In the 335 \AA~ image only the leading edges are
discernible. However, a loop from the umbra at $x=6\arcsec$
and $y=13\arcsec$ is hinted. The visibility of the leading edges and
umbral loops above the LB is similar to that in the 171 \AA~,
211 \AA~ and in the 193 \AA~ image. The bright part of the coronal
loops seen in the umbra are plumes (Foukal 1978). A detailed
analysis of plumes has been presented by Kleint et al. (2014) using
IRIS spectra.  The structuring of the LB in the 1700 \AA~ image
is similar to that in Ca II H but due to the lower resolution of
AIA fine structures are not clear. Even bright threads seen in the
G-band and Ca II H are not visible. The bright threads as well as the
bright leading edges are also visible in the 1600 \AA~ image. Berger
\& Berdyugina (2003) reported enhanced contrast above a LB structure
in the 1600 \AA~ images taken from the TRACE. The resolution of their
data was not high enough to resolve the fine scale details observed by
IRIS.

\section{Discussion and conclusion}

With high spatial and temporal resolution slit jaw images from the
IRIS at 1330 \AA~ we are able to demonstrate fine scale details of
jets above a LB in the TR which were hidden in earlier
observations from the ground and from space.  Particularly even at the
double resolution of Ca II H images from Hinode such fine
details and structuring have not been seen. Louis et al
(2004) presented a detailed study of chromospheric jets above a LB
that show triangular shaped blobs. The top part of the jets extend
into a spike-like structure. Thus such spikes have a bright base and
the top part is extended into the dark umbral background with
decreasing brightness toward outer edge.  This is the general
brightness structure of the Ca II H jets. However, some jets show the
bright leading edge as well.  The main difference found in the present
analysis belongs to the 1330 \AA~ images. In most of the
cases only the bright leading edge of the jets are seen in the dark
umbral background with high contrast. The section of jets between the
apex and the base is almost void in the 1330 \AA~ while no such void
is seen in the Ca II H images.  Only very bright jets show continuous
structure in the 1330 \AA~ similar to the Ca II H.

Jets, blobs or brightening seem to be essential parts of the chromosphere above LBs.
However, Bharti et al. (2015 in preparation) suggested that such activity is preferentially seen above filamentary LBs.
Lagg et al. (2014) analysed a granular LB and did not find such activity in the Ca II H images. Low altitude reconnection
has been suggested as mechanism to produce such an activity (See Bharti et al. 2007 and reference their in).
According to recent MHD simulations,
sunspot fine structures are caused my magnetoconvection. Cheung et al. (2010) has shown that simulated
LBs have upflows in the central part and downflows at the lateral edges similar to penumbral filaments.
These strong downflows at the lateral edges drags field lines and forms a hairpin like structure with opposite
polarity (Bharti et al. 2010, Rempel et al. 2011). Such opposite polarity fields have been reported in penumbral filaments by
Franz \& Schlichenmaier (1013) and Scharmer et al. (2013). With new inversion schemes where the PSF of the telescope has been taken
into account, more opposite polarity fields are reported (Ruiz Cobo \& Asensio
  Remos 2013, van Noort 2012). Using synthetic spectra from a simulated sunspot Bharti et al. (2015) confirmed
this scenario. Bharti et al. (2015b) find a significant amount of opposite polarity field at the lateral edges
of LBs using spatially coupled inversion (van Noort 2012) on Hinode/SP data that is accompanied with brightening in Ca II H.
 Louis et al. (2014) also reported isolated patches of opposite polarity around the
location of jets observed in Ca II H . Bharti et al. (2015a) reported $\lambda$-shaped jets
above a LB in the Ca II H images observed at the SST and opposite polarity patches in the photospheric
magnetic field at the edge of the LB. Thus reconnection occurs between a newly emerged opposite polarity
patch and the preexisting umbral field. Such a reconnection scenario is suggested to explain jets (or $\lambda$-shaped jets)
seen in the chromosphere (Yokoyama \& Shibata 1995, Shibata et al. 2007).
Thus triangular shaped jets reported by Louis at al. (2014) could be $\lambda$-shaped and jets reported
by Bharti et al. (2015a) could be more clear due to the higher resolution of the SST data and the advantageous viewing angle.
We also clearly see the effect of the viewing angle on the data presented here as jets are seen only on the right side
(opposite side of disk center) of the LB, laying over the umbral background. There are hints of $\lambda$-shaped jets
but viewing angle and resolution is not ideal to see them distinctly. The spatial and temporal resolution of IRIS data is
good enough to detect small scale dynamics such as mass motion and loop interaction which was not possible from
Hinode Ca II H data (Louis et al. 2014). Using the Ca II H Hinode/SOT observations Shimitzu et al. (2009) and Shimitzu (2011)
reported about intermittent plasma ejections above a LB for a few days, accompanied with a change in the photospheric magnetic
flux density and inclination. Shimizu (2009) proposed a model where the LB is considered as a highly twisted
current-carrying flux tube lieing below the background field that forms a cusp-like shape above the LB. The proposed geometry
then produces opposite polarity field at one side of LB and led to reconnection. Numerical simulations
by Magara (2010) and Nishizuka et al. (2012) succeeded to produce jet-like structures in a penumbral configuration
(Katsukawa et al. 2007) where reconnection occurs between the more horizontal penumbral and more vertical background field.
This scenario is similar in the case of a LB having horizontal field and slightly inclined umbral field (Jur\v c\'ak et al. 2006).

The main findings of the current study are enhanced emission along the
leading edge of jets and coordinated behavior between neighbouring
jets observed in the IRIS and AIA wavebands. Rouppe van der Voort \&
de la Cruz Rodr\'iguez 2013) find jets (dynamic fibrils with parabolic
path) whose properties vary from the sunspot umbra to beyond the
penumbra. This reinforces the role of convective
energy and inclination on the propagation of long period waves that
form shocks in the chromosphere. However, these jets are found to be
different from penumbral micro-jets (Katsukawa et al. 2007) and umbral
micro-jets (Bharti et al. 2013) which are caused by reconnection. The
LB jets studied here are caused by reconnection and we speculate that
a wave phenomenon (leakage of waves through the LB)
is responsible for the coordinated
behavior between neighbouring jets.  Morton (2012) reported on jets at
the edge of a sunspot in Ca II H Hinode observations and
suggested that the enhanced emission is caused by shock heating.  Thus
plasma is heated by a shock front at TR temperatures around
the leading edge of the jets and only the leading bright edge is
visible in most of the LB jets. However, we cannot rule out other
mechanisms and topologies that can just compress the jet
material, i.e., increases its density, which, under optically
thin conditions, leads to increased emission (Morton 2015, private communication).
The remaining part of
the jets remain cooler in the TR. Some jets in the Ca II H images show
an entire bright structure with the leading edge which indicates that jets are at
chromospheric temperature. The coronal counterpart of the jets also
show the same behaviour, as evident from the 211 \AA~ and 193
\AA~images that only the leading edges appear brighter.  Thus the
jet's leading edges are heated up to coronal temperatures. However, these channels also
contain significant TR contribution, which may well be the cause for the observed brightness.  In
the TR the most part of the jet is as cool as the ambient
umbra so the it is hardly visible while the leading edges are at TR
temperatures, thus they appear bright. The lack of visibility of the body of the jets in the 1330 \AA~ passband
may be caused by the fact that the passband contains both FUV continuum that is formed at lower chromospheric temperatures (likely
the reason of the bright bottom of the jets in C II) as well as the C II lines which are formed at the upper chromospheric/low TR
temperatures, thus the leading edges appear brighter. The ejected blobs of plasma
seen bright in the TR and coronal wavelength bands are also sources of
mass and heat. This suggests that such jets are important
media to transfer energy from the photosphere to the corona via
the TR above the sunspot.

Spectral analysis of LB jets and their leading edge from the IRIS
spectra in various wavelengths will be useful to understand various
properties of these small scale dynamic jets (Bharti et al. in
preparation).

\section*{Acknowledgments}

The author thanks anonymous referee for useful criticism to improve the presentation of this manuscript.
Dr. J. Hirzberger gratefully acknowledged for carefully reading the manuscript.
IRIS is a NASA small explorer mission developed and operated by LMSAL with mission operations executed at NASA
Ames Research center and major contributions to downlink communications funded by the Norwegian Space Center
(NSC, Norway) through an ESA PRODEX contract. Hinode is a Japanese mission developed and launched by
ISAS/JAXA, collaborating with NAOJ as a domestic partner and NASA and
STFC (UK) as international partners. Scientific operation of the Hinode mission
is conducted by the Hinode science team organized at ISAS/JAXA. This team
mainly consists of scientists from institutes in the partner countries. Support for
the postlaunch operation is provided by JAXA and NAOJ (Japan), STFC (UK), NASA, ESA, and NSC (Norway).

\end{document}